\documentclass[12pt]{article}
\usepackage{graphicx}

\def\slash#1{\rlap{\hbox{$\mskip 1 mu /$}}#1}      
\def\Slash#1{\rlap{\hbox{$\mskip 3 mu /$}}#1}      
\def\dg{\dagger}
\def\ca{{\cal A}}

\def\c{\chi}
\def\d{\delta}
\def\e{\epsilon}
\def\f{\phi}
\def\g{\gamma}
\def\h{\eta}

\def\j{\psi}

\def\l{\lambda}
\def\m{\mu}
\def\n{\nu}

\def\p{\pi}

\def\r{\rho}
\def\s{\sigma}

\def\x{\xi}

\def\D{\Delta}

\def\G{\Gamma}

\def\L{\Lambda}
\def\O{\Omega}

\def\S{\Sigma}

\def\R{\mathcal R}

\def\beq{\begin{equation}}
\def\eeq{\end{equation}}
\def\bea{\begin{eqnarray}}
\def\eea{\end{eqnarray}}

\def\NO{\nonumber}

\def\pl#1#2#3{Phys.~Lett.~{\bf B {#1}} ({#2}) #3}
\def\np#1#2#3{Nucl.~Phys.~{\bf B {#1}} ({#2}) #3}
\def\prl#1#2#3{Phys.~Rev.~Lett.~{\bf #1} ({#2}) #3}
\def\pr#1#2#3{Phys.~Rev.~{\bf D {#1}} ({#2}) #3}

\def\ap#1#2#3{Ann.~of Phys.~{\bf {#1}} ({#2}) #3}

\def\ptp#1#2#3{Progr.~Theor.~Phys.~{\bf {#1}} ({#2}) #3}

\renewcommand{\(}{\left(}
\renewcommand{\)}{\right)}
\renewcommand{\[}{\left[}
\renewcommand{\]}{\right]}

\begin{document}
\date{\mbox{ }}
\title{{\normalsize DESY 02-115\hfill\mbox{}\\
September 2002\hfill\mbox{}}\\
\vspace{2cm} \textbf{Bulk and Brane Anomalies\\ In Six Dimensions}\\
[8mm]}
\author{T.~Asaka, W.~Buchm\"uller, L.~Covi\\
\textit{Deutsches Elektronen-Synchrotron DESY, Hamburg, Germany}}
\maketitle

\thispagestyle{empty}

\begin{abstract}
\noindent
We study anomalies of six-dimensional gauge theories compactified on
orbifolds. In addition to the known bulk anomalies, brane anomalies appear
on orbifold fixpoints in the case of chiral boundary conditions. At a
fixpoint, where the bulk gauge group G is broken to a subgroup H, 
the non-abelian G-anomaly in the bulk reduces to a H-anomaly which 
depends in a simple manner on the chiral boundary conditions. 
We illustrate this mechanism by means of a SO(10) GUT model.
 
\end{abstract}

\newpage

\section{Introduction}

The structure of the standard model of strong and electroweak interactions, 
its gauge group and field content, 
points towards an underlying unified theory (GUT) of all particles and 
interactions. The simplest GUT group which unifies the gauge interactions of the
standard model is SU(5) \cite{gg74}. With the present evidence for neutrino masses
and mixings the larger gauge group SO(10)~\cite{gfm75} appears particularly 
attractive. It contains SU(5) as well as the Pati-Salam group 
SU(4)$\times$SU(2)$\times$SU(2)~\cite{ps74} 
and flipped SU(5)~\cite{fl82} as subgroups.
 
The quest for unification with gravity points towards supersymmetry and higher 
dimensions. Orbifold compactifications \cite{dhx85} then provide a promising bridge 
to the four-dimensional world since they generically lead to chiral gauge theories
as effective theories in lower dimensions. Hence, orbifold compactifications provide 
an attractive starting point for attempts to embed the standard model of particle 
physics into higher dimensional string and field theories.

Orbifold compactifications also allow to break the gauge symmetry of grand unified
theories to the standard model gauge group
in an attractive and simple manner. In particular, the breaking of the GUT 
symmetry automatically yields the required doublet-triplet splitting of Higgs fields 
\cite{kaw00}. Several SU(5) models have been constructed in five dimensions (5d) 
\cite{kaw00}-\cite{hmr01}, whereas six dimensions are required for the 
breaking of SO(10) \cite{abc01,hnx02}. Global anomaly cancellation \cite{dp01}
or extended supersymmetry \cite{wy02} in 6d can also be used to explain the number 
of quark-lepton generations.

In general, orbifold compactifications lead to anomalies at orbifold fixpoints.
So far, this has been studied for U(1) symmetries in 5d theories 
\cite{acg01}-\cite{gno02} and for 10d heterotic orbifolds \cite{ggnow02}
, where no bulk anomalies exist. The cancellation of
the brane anomalies at orbifold fixpoints is crucial for the consistency
of the orbifold compactification and the field content of the theory.

In the present paper we investigate anomalies in orbifold
compactifications of 6d theories. This is motivated by recently proposed
supersymmetric 6d GUT models. Contrary to five dimensions, bulk anomalies 
exist in six dimensions for N=1 supersymmetry, and the
question arises how brane and bulk anomalies are related.

It turns out that Fujikawa's method of calculating anomalies is particularly
well suited to study this question. In section~2 we shall explicitly calculate the
U(1) anomaly of a 6d Weyl fermion on the orbifold $M=\R^4\times T^2/Z_2$
and compare the result with the anomaly in flat space $M=\R^6 $ and
on the torus, $M=\R^4\times T^2$. 
In section~3 we  extend this result to non-abelian anomalies
and determine the general connection between the brane anomalies and the chiral
boundary conditions at orbifold fixpoints. This pattern will be illustrated in
more detail in section~4 by means of the SO(10) GUT model proposed in \cite{abc021}.
Our results are summarized in section~5, and some useful formulae are collected
in the appendices.

\section{The abelian anomaly in six dimensions}

Consider a Weyl fermion $\j$ with U(1) gauge interaction in six dimensions, 
which is described by the lagrangian
\bea
{\cal L} = \bar{\j}(z) i\G^M D_M \j(z) \;.
\eea
Here $D_M = \partial_M + A_M$, $M=1\ldots 6$, is the covariant derivative with 
field strength $F_{MN} = [D_M,D_N]$\footnote{Our conventions for the $\G$-matrices 
are listed in appendix~A.}. The 6d Weyl fermion is composed of two 4d Weyl fermions 
with opposite 4d chirality, $\j = (\j_L,\j_R)$, with $\g_5 \j_L = -\j_L$ and 
$\g_5 \j_R = \j_R$; $\j $ has negative 6d chirality, i.e. $\G_7 \j = -\j$, where 
$\G_7$ = diag($\g_5,-\g_5$).

Naive dimensional reduction to five dimensions yields a U(1) gauge theory 
with a Dirac fermion, $\c = \j_L + \j_R$, with U(1) gauge interaction,
\bea
{\cal L} = \bar{\c}(z) i\g^M D_M \c(z) \;,
\eea
where $\g^M$, $M=1\ldots 5$, are the usual 4d $\g$-matrices. This model has been
discussed in the literature in connection with anomalies arising 
on the orbifold $S^1/Z_2$ \cite{acg01}-\cite{gno02}.
 
We now consider the compactification of the 6d theory on the 
orbifold $M=\R^4\times T^2/Z_2$. The two elements of the group $Z_2$ are the
identity and the reflection at one point on the torus $T^2$, e.g.
$y\rightarrow -y$, where $y = (z^5,z^6)$. The orbifold $T^2/Z_2$ has four fixpoints,
$y_1 = (0,0)$, $y_2 = (\p R_5,0)$, $y_3 = (0,\p R_6)$ and $y_4 = (\p R_5,\p R_6)$,
which correspond to the four corners of a `pillow'. Here $R_5,R_6$ are the radii of
the torus in the $z^5$ and $z^6$ direction respectively.
For the fermion $\j$ we
impose chiral boundary conditions, 
\bea\label{chiral}
\j_L(x,y) = \j_L(x,-y)\;, \quad \j_R(x,y) = -\j_R(x,-y)\;,
\eea
where $x$ denotes the coordinates of flat 4d Minkowski space. 
In terms of the complete
system of mode functions (cf. appendix~\ref{chap:appendixc}), 
the fermions $\j_L$ and $\j_R$ can
be expanded as
\bea
\j_L(x,y) = \sum_{mn} \j_{L+}^{mn}(x) \x^{mn}_+(y)\; , \quad
\j_R(x,y) = \sum_{mn} \j_{R-}^{mn}(x) \x^{mn}_-(y)\; .
\eea
Invariance of the lagrangian under the $Z_2$ symmetry requires for the background
gauge field,
\bea
A_\m(x,y) = A_\m(x,-y)\;,\quad 
A_{5,6}(x,y) =-A_{5,6}(x,-y)\;.
\eea
Note, that $A_{5,6}$ vanishes at the fixpoints $y_i$, $i=1\ldots 4$. 

The effective action $\G[A]$, which is defined by
\bea
e^{i\G[A]} = \int D\j D\bar{\j} \exp\(i\int d^6z {\cal L}\)\;,
\eea
transforms under infinitesimal gauge transformations 
$\d_v A_M = \partial_M v$ as
\bea
\d_v \G[A] = \int d^6z\, \left( \partial^M \left[ v(z)\, J_M(z) \right] 
- v(z) \partial^M\! J_M(z) \right) \; ,
\label{abel-variation}
\eea
where $J_M(z) = \d \G[A]/\d A^M(z)$ is the U(1) current.
We have kept for generality the boundary term due to the partial 
integration. In the case of singular currents and manifolds with
boundaries, like in the orbifold case, a contribution from 
the boundary can survive \cite{ch85}.
Due to the non-invariance of the measure $D\j D\bar{\j} $ gauge
invariance is spoiled \cite{fuj79},
\bea
\d_v \G[A] = - \int d^6z\, v(z) \ca(z) \; .
\label{abel-anom}
\eea
For vanishing boundary term the divergence of the current is then given by 
the anomaly \cite{abj69},
\bea
\partial^M\!J_M(z) = \ca(z)\; ,
\eea
which can be expressed as a trace over modes of
$\j$ and $\bar{\j}$, respectively \cite{fuj79}. 

Let $\f_n$ be a complete set of
eigenfunctions $\f_n$ of the hermitian operator ${\Slash D}^2 = (\Gamma^M D_M)^2$ 
with eigenvalues $\l_n^2$, i.e.  ${\Slash D}^2 \f_n = \l_n^2\f_n$. 
A left-handed 6d Weyl fermion 
$\j$ can be expanded into eigenfunctions of ${\Slash D}^2$ and $(1-\G_7)/2$.
Correspondingly, $\bar{\j}$ is right-handed and can be expanded in eigenfunctions
of ${\Slash D}^2$ and $(1+\G_7)/2$. The anomaly is then given by the 
difference of sums over left-handed and right-handed modes, respectively 
\cite{fuj79,zwz84,agg85},
\bea
\ca(z) =  \lim_{\L\to\infty} \sum_n \(\f_n^\dg(z){1-\G_7\over 2} \f_n(z) 
                  -\f_n^\dg(z){1+\G_7\over 2} \f_n(z) \) e^{-\l_n^2/\L^2}\;,
\eea
where the sum has been regularized by the ultraviolet cutoff $\L$.
Choosing plane waves as eigenfunctions in flat space, one obtains \cite{zwz84},
\bea\label{flat}
\ca(z) 
&=& -\lim_{\L\to\infty} {\rm Tr} \int {d^6k\over (2\p)^6} \G_7
         e^{ikz} e^{-{\slash D}^2/\L^2} e^{-ikz} \NO\\
&=& -\lim_{\L\to\infty} {\rm Tr} \int {d^6k\over (2\p)^6} \G_7
         \exp\( {(k+i D)^2\over \L^2} -  {1\over 4\L^2}[\G^M,\G^N]F_{MN}\) \NO\\
&=& -\lim_{\L\to\infty} {1\over 3!} {\rm Tr}\ \G_7 \({-1\over 4\L^2}[\G^M,\G^N]F_{MN}\)^3
    \L^6\, \int {d^6k\over (2\p)^6} e^{k^2} \NO\\
&=& - \,{i^3\over 3!(4\p)^3} \e^{MNPQRS} F_{MN} F_{PQ} F_{RS}\;.    
\eea
Here ${\rm Tr}$ denotes the trace over Dirac matrices in 6d, and
after Wick rotation to Euclidean space the metric is $\h_{MN}^E=-\d_{MN}$. 

If two of the six dimensions are compactified on a torus one can choose as 
eigenfunctions the product of 4d plane waves with the orthonormal modes 
$\x^{mn}_\pm$ on $T^2$ (cf. appendix~\ref{chap:appendixc}). The sum over all modes then reads
\bea
{\rm Tr} \int {d^4k\over (2\p)^4} e^{k^2/\L^2} 
         \sum_{mn} e^{-{M_m^2+M_n^2\over \L^2}} \({\x_+^{mn}}^2(y) 
+ {\x_-^{mn}}^2(y)\) \;,
\label{torus-anom}
\eea
which, in the limit $\L R_{5,6}\rightarrow \infty$, becomes the 6d sum of flat space,
i.e. $\int d^6k/ (2\p)^6 \exp{(k^2/\L^2)}$. Hence, the abelian anomaly on
$M=\R^4\times T^2$ is identical to the one in flat space.
  
Consider now compactification on the orbifold $M=\R^4\times T^2/Z_2$. 
In this case the physical space corresponds to the pillow with corners 
$y_1 = (0,0)$, $y_2 = (\p R_5,0)$, $y_3 = (0,\p R_6)$ and 
$y_4 = (\p R_5,\p R_6)$, with half the volume of the torus.
The variation of the action then reads
\bea
\d_v \G[A] &=& - \int d^4x\,\int_{T^2/Z_2}\! d^2y\, 
v(x,y) \partial^M\! J_M(x,y) \\
&=&  - \int d^4x\,\int_{T^2/Z_2}\! d^2y\, v(x,y) \ca (x,y)\; \\
&=&  - \int d^4x\,\int_{T^2}\! d^2y\, v(x,y) \ca_{cov} (x,y)\; ,
\label{abel-anom-orbi}
\eea
where in the last line we have extended the integral to
the covering space $T^2$. In this way we can resort to the trick of
using mode functions on $T^2$ and compare more directly the
result with the torus case. For the relation between $\ca $ and
$\ca_{cov}$ see appendix D.

Another difference is that on the orbifold the chiral boundary 
conditions (\ref{chiral}) have
to be taken into account in the sum over the modes of $\j$ and $\bar{\j}$. 
This can be done by means of the projection operators
\bea
{1\pm \G_7\over 2} \hat{P}_{L(R)}\;,
\label{4d-project}
\eea
where the 4d chirality operator acting on 6d spinors is defined as
\bea
\hat{P}_{L(R)} = 
\(\begin{tabular}{cc} $P_{L(R)}$ & 0 \\ 0 & $P_{L(R)}$  \end{tabular}\)\;,
\eea
and $P_{L(R)} = (1\mp \g_5)/2$ is the usual 4d chiral projector. 
The operators in eq.~(\ref{4d-project}) single out the components 
$\j_{L(R)}$ of the 6d Weyl spinor $\j$. 
For the anomaly one then obtains (cf.~(\ref{flat})),
\pagebreak
\bea\label{orbi6}
\ca_{cov}(x,y) 
&=& \lim_{\L\to\infty} {\rm Tr} \int {d^4k\over (2\p)^4} e^{ikx} 
\sum_{mn} e^{-{\slash D}^2/\L^2} e^{-ikx} \\
&& \hspace{0.6cm} \times \left[{1-\G_7\over 2}\(\hat{P}_L {\x_+^{mn}}^2(y) 
+ \hat{P}_R {\x_-^{mn}}^2(y)\) \right. \NO\\
&&\hspace{1cm} \left.
- {1+\G_7\over 2}\(\hat{P}_R {\x_+^{mn}}^2(y) 
+ \hat{P}_L {\x_-^{mn}}^2(y)\)\right] ,\NO 
\eea
which is conveniently expressed as
\bea
\ca_{cov}(x,y)\label{corbi6} 
&=& - {1\over 2}\lim_{\L\to\infty} {\rm Tr} \int {d^4k\over (2\p)^4} e^{k^2/\L^2}
\sum_{mn} e^{-{M_m^2+M_n^2\over \L^2}} \exp{\({-1\over 4\L^2}[\G^M,\G^N]F_{MN}\)}\\
&&\hspace{1.2cm} \times\left[\G_7\({\x_+^{mn}}^2(y) + {\x_-^{mn}}^2(y)\) 
   + \(\hat{P}_R - \hat{P}_L\) \({\x_+^{mn}}^2(y) - {\x_-^{mn}}^2(y)\)\right]\ .\NO
\eea
The term proportional to $(\x_+^2+\x_-^2)$ is identical to the anomaly
on the torus, up to a factor $1/2$.  Hence, we obtain on the covering
space half the bulk anomaly of flat space. This is plausible since we
have projected out half of the modes.  In fact we can write the torus
wavefunction as a sum of two orbifold wavefunctions with opposite
parities and recover the result of eq.~(\ref{torus-anom}). Remember
anyway that the orbifold bulk anomaly on the physical space is larger by a factor 2 
(cf. appendix~\ref{chap:appendixd}), so that locally one cannot
distinguish the global properties of the space.

On the other hand, the sum over the difference of modes, 
$(\x_+^2-\x_-^2)$, is finite (cf. appendix~\ref{chap:appendixc}), and independent
of the cut-off,
\bea
\sum_{mn}\({\x_+^{mn}}^2(y) - {\x_-^{mn}}^2(y)\) = \d_O(y) \;.
\eea
Correspondingly, taking the limit $\Lambda \rightarrow \infty$,
the term proportional to 
${\rm Tr}\(\hat{P}_R - \hat{P}_L\)\([\G^M,\G^N]F_{MN}\)^3$ vanishes, 
whereas  a term
${\rm Tr}\(\hat{P}_R - \hat{P}_L\)\([\G^M,\G^N]F_{MN}\)^2$ 
survives, proportional
to the 4d anomaly. Combining both terms we finally obtain for the anomaly,
\bea\label{aan-cov}
\ca_{cov} (x,y) = 
- \,{1\over 2} {i^3\over 3!(4\p)^3} \e^{MNPQRS} F_{MN} F_{PQ} F_{RS}
+ {i^2\over 2!(4\p)^2} \d_O(y) \e^{\m\n\r\s} F_{\m\n} F_{\r\s}\;.
\eea
As described in the appendix~\ref{chap:appendixd}, the anomaly on the physical
space $T^2/Z_2$ reads then
\bea\label{aan}
\ca (x,y) = -\, {i^3\over 3!(4\p)^3} \e^{MNPQRS} F_{MN} F_{PQ} F_{RS}
+ {i^2\over 2!(4\p)^2} \d_O(y) \e^{\m\n\r\s} F_{\m\n} F_{\r\s}\;.
\eea

The interpretation of this result is obvious: the first term is 
the usual 6d bulk anomaly, and the second term,
generated by the chiral boundary conditions at the orbifold fixpoints, 
is a localized 4d anomaly. 
Note that the sum of the 4d anomalies at the fixpoints equals the 4d anomaly 
of the zero mode $\j_L^{00}$.
In fact the contributions of the massive modes to the integrated anomaly
compensate each other for every Kaluza-Klein level $(m,n)$.
In the effective 4d low energy theory therefore only the contribution
of the zero modes survives, if the bulk anomaly vanishes.

For comparison, it is instructive to compute also the abelian anomaly in 
five dimensions, on the orbifold $M=\R^4\times S^1/Z_2$. The two fixpoints are 
$y_1 = 0$ and $y_2 = \p R_5$, with $y = z^5$.
The chiral boundary conditions are again given by eq.~(\ref{chiral}).
Fermions are now four-component spinors, $\chi = \j_L + \j_R$, and left- and
right-handed spinors can be expanded in terms of $\x^m_+$ and $\x^m_-$, respectively
(cf. appendix~\ref{chap:appendixc}).
The trace formula (\ref{orbi6}) for the 6d anomaly then becomes
\bea\label{orbi5}
\ca_{cov} (x,y) 
&=& \lim_{\L\to\infty} {\rm Tr} \int {d^4k\over (2\p)^4} e^{ikx} 
\sum_{m} e^{-{\slash D}^2/\L^2} e^{-ikx} \NO\\
&& \hspace{1.2cm} \times \left[\(P_L {\x_+^{m}}^2(y) + P_R {\x_-^{m}}^2(y)\)
- \(P_R {\x_+^{m}}^2(y) + P_L {\x_-^{m}}^2(y)\)\right]\; , 
\eea
which yields
\bea
\ca_{cov} (x,y) 
&=& - \lim_{\L\to\infty} {\rm Tr} \int {d^4k\over (2\p)^4} e^{k^2/\L^2}
\sum_{m} e^{-{M_m^2\over \L^2}} \exp{\({-1\over 4\L^2}[\G^M,\G^N]F_{MN}\)} \NO\\
&&\hspace{2.6cm} \times\left[\g_5 \({\x_+^{m}}^2(y) - {\x_-^{m}}^2(y)\)\right]\ .
\eea
As on the torus, the sum over the differences of modes, $(\x_+^2-\x_-^2)$, is
finite, and one finally obtains
\bea
\ca_{cov} (x,y) = \ca (x,y) = {1\over 2} \left(\d(y) + \d(y-\p R^5)\right)
{ i^2\over 2!(4\p)^2} \e^{\m\n\r\s} F_{\m\n} F_{\r\s}\;.
\eea
This result has previously been obtained \cite{acg01} by direct evaluation of the 
divergence of the 5d U(1) current, using the known 4d anomaly, and also by means of
Fujikawa's method \cite{bcx02}.

\section{The non-abelian anomaly}

The abelian anomaly (\ref{aan}) is most conveniently written as differential form. With
\bea
A = A_M dz^M\;, \quad  F = d A = {1\over 2} F_{MN} dz^M dz^N\;,
\eea
one obtains for the 6-form $\hat{\ca} = \ca(z) dz^1\ldots dz^6$,
\bea
\hat{\ca} = - \,{i^3\over (2\p)^3} F^3 
            + \d_O(y)dz^5 dz^6 {i^2\over (2\p)^2} F^2\;,
\eea
where wedge products are understood. 

Consider now a 6d Weyl fermion $\j$ in a non-abelian background field which is an
element of the Lie algebra, i.e. $D_M = \partial_M + A_M$ and $A_M = i A_M^a T^a$,
where $T^a$ are the generators of the group G. Field strength and gauge variation
are now
\bea
F = d A + A^2\;, \quad \d_v A = d v + [A,v] \;,
\eea
where $v = i v^a T^a$. The variation of the effective action, neglecting the
boundary term, is given by
\bea
\d_v \G[A] = - \int d^6z\, v^a(z)\(\ca^a(z) + \D_{WZ}^a(z)\)\; . 
\eea
The non-abelian anomaly $\ca^a + \D_{WZ}^a$ satisfies the Wess-Zumino consistency 
conditions \cite{wz71}. 
It differs from the covariant anomaly $\ca^a$ by $\D_{WZ}^a$, a local 
polynomial in the gauge field \cite{agg85}. Since we are only interested in the
question of anomaly cancellation, we can ignore this difference and consider just
the covariant anomaly which is again given by a trace formula \cite{agg85},  
\bea
\ca^a(z) =  \lim_{\L\to\infty} \sum_n \(\f_n^\dg(z) T^a{1-\G_7\over 2} \f_n(z) 
                  -\f_n^\dg(z) T^a {1+\G_7\over 2} \f_n(z) \) e^{-\l_n^2/\L^2}\;.
\eea
A calculation completely analogous to the one in section~2 then yields for the
non-abelian anomaly on the orbifold $\R^4\times T^2/Z_2$,
\bea\label{nonab}
\hat{\ca}^a (x,y) = -\; {i^3\over (2\p)^3} {\rm tr}\(T^a F^3\) \;
          +\; \d_O(y)dz^5 dz^6 {i^2\over (2\p)^2} {\rm tr}\(T^a F^2\)\;,
\eea
where ${\rm tr}$ denotes the trace over the fermion representation of the group G.

Boundary conditions at orbifold fixpoints can be used to break the group G to a
symmetric subgroup H. This is achieved by means of an automorphism of the Lie 
algebra, characterized by a parity operator $P$, with $P^2 = I$.
For the gauge field $A$, the corresponding boundary conditions read
\bea
P A_\m(x,-y) P^{-1} &=& +A_\m(x,y)\;, \quad 
P A_{5,6}(x,-y) P^{-1} = -A_{5,6}(x,y)\; .
\eea 
Note, that $P$ acts differently on the generators $T^{\tilde a}$ of $H$ 
and $T^{\hat a}$ of $G/H$,
\bea
P T^{\tilde a} P^{-1} = + T^{\tilde a} \;  , \quad
P T^{\hat a} P^{-1} &=& - T^{\hat a}\; ,
\eea 
allowing zero modes only for $A^{\tilde a}_\mu $ and $A^{\hat a}_{5,6}$. 
Also the 6d gauge transformations are restricted to 
those with $ \partial_\mu v^{\hat a} (x,0) = 0\, ,
\partial_{5,6} v^ {\tilde a}(x,0) = 0$. 
Hence, only the local symmetry corresponding to $H$ is present
at the orbifold fixed point.

The 6d Weyl fermion, $\j = (\j_{L},\j_{R})$, splits into two,
in general reducible, representations of H,  $\j = (\j_{1},\j_{2})$,
which have positive and negative parity, respectively,
\bea
P\j_{1}(x,y) = + \j_{1}(x,y)\;, \quad 
P\j_{2}(x,y) = - \j_{2}(x,y)\; .
\eea
The chiral boundary condition (\ref{chiral}) then becomes
\bea
P\j_{L1}(x,-y) &=& + \j_{L1}(x,y)\;, \quad P\j_{L2}(x,-y) = - \j_{L2}(x,y)\;,
\label{chiralP1}\\
P\j_{R1}(x,-y) &=& - \j_{R1}(x,y)\;, \quad P\j_{R2}(x,-y) = + \j_{R2}(x,y)\;.
\label{chiralP2}
\eea
These boundary conditions allow only two 4d zero modes, one left- and 
one right-handed fermion in two different representations of H, 
which can be characterized by the projection operators
$P_{1} = (1 + P)/2$ and $P_{2} = (1 - P)/2$. 

We can now again calculate the non-abelian anomaly on the orbifold with the
new boundary conditions which break G to H. The anomaly is given by the same 
expression as (\ref{orbi6}) except for the mode sum which has to be replaced
by
\bea\label{Porbi6}
&&\sum_{mn} e^{-{M_m^2+M_n^2\over \L^2}}  \left\{ {1-\G_7\over 2}\left[ 
\left(\hat P_L P_1 + \hat P_R P_2 \right) {\x_+^{mn}}^2(y) 
+ \left(\hat P_L P_2 + \hat P_R P_1 \right) {\x_-^{mn}}^2(y)
\right]\right. \NO\\
&&\hspace{0.8cm}\left.-{1+\G_7\over 2}\left[ 
\left(\hat P_R P_1 + \hat P_L P_2 \right) {\x_+^{mn}}^2(y) 
+ \left(\hat P_R P_2 + \hat P_L P_1 \right) 
{\x_-^{mn}}^2(y)\right]\right\}\;. 
\eea
This expression can again conveniently be written in the form 
of eq.~(\ref{corbi6}), with the mode sum,
\bea
&& \sum_{mn} e^{-{M_m^2+M_n^2\over \L^2}} 
\left\{\G_7\({\x_+^{mn}}^2(y) + {\x_-^{mn}}^2(y)\)\right. \NO\\ 
&&\hspace{1.5cm}\left.+ \(\hat{P}_R - \hat{P}_L\)\(P_1-P_2\) 
  \({\x_+^{mn}}^2(y) - {\x_-^{mn}}^2(y)\)\right\}
\label{chiralP}\; .
\eea
Note that, as before, $\hat{P}_R - \hat{P}_L =$  diag$(\gamma_5, \gamma_5) $,
while  $ P_1-P_2 = P $.

The final expression for the anomaly then reads
\bea
\hat{\ca}^a (x,y) &=& - \, {i^3\over (2\p)^3} {\rm tr}\(T^a F^3\) 
          + \d_O(y)dz^5 dz^6 {i^2\over (2\p)^2} {\rm tr}\((P_1-P_2)T^a F^2\)\\
&=& - \,{i^3\over (2\p)^3} {\rm tr}\(T^a F^3\) 
          +\d_O(y)dz^5 dz^6 {i^2\over (2\p)^2} {\rm tr}\(P T^a F^2\)\;.
\eea
The only difference with respect to eq.~(\ref{nonab}), the anomaly in the 
case without symmetry breaking, is the appearance of projection operators,
and therefore of the parity operator $P$, in the second term. 
At the fixpoint, the group G is broken to the subgroup H. 
It is therefore consistent to have in the fixpoint term of the anomaly 
projection operators $P_1$ and $P_2$ for the two different representations 
of H. The relative sign is different, since the chiral boundary conditions 
(\ref{chiralP1}), (\ref{chiralP2}) associate a 4d left-handed 
fermion with $P_1$ and a 4d right-handed fermion with $P_2$. 

At the fixpoint only the gauge group $H$ can act, and the gauge 
variation $\partial_\mu v^{\hat a}$ for the coset $G/H$ vanishes there. 
Correspondingly, for the localized anomaly the trace 
${\rm tr}\(P T^{\hat a} F^2\) $
vanishes for any generator $T^{\hat a}$ belonging to the coset $G/H$, 
since we have
\bea
{\rm tr}\(P T^{\hat a} F^2\) = - {\rm tr}\(P T^{\hat a} F^2\) = 0\; ,
\eea
from $ P T^{\hat a} = - T^{\hat a} P $ and $ P F^2 = F^2 P $.

Similarly, also the bulk anomaly at the fixpoint is non-zero only for 
generators $T^{\tilde a}$ belonging to $H$. 
In fact, there the non-vanishing fields 
are $F^{\tilde a}_{\mu\nu}, F^{\tilde a}_{56} $ and 
$ F^{\hat a}_{\mu 5}, F^{\hat a}_{\mu 6} $. Hence, the only completely 
antisymmetric terms are of the type
\bea
{\rm tr}\(T^a T^{\tilde a} T^{\tilde a'} T^ {\tilde a''}\) 
\epsilon^{\mu\nu\rho\sigma}  F^{\tilde a}_{\mu\nu} 
F^{\tilde a'}_{\rho\sigma}  F^{\tilde a''}_{56}\; ,
\eea
corresponding to the bulk $H$ anomaly term, and the mixed piece
\bea
{\rm tr}\(T^a T^{\tilde a} T^{\hat a} T^{\hat a'} \) 
\epsilon^{\mu\nu\rho\sigma}  F^{\tilde a}_{\mu\nu} F^{\hat a}_{\rho 5}  
F^{\hat a'}_{\sigma 6}\; .
\label{mix-anomaly}
\eea
Both group traces vanish identically for generators $T^a$ belonging to
$G/H$, since they contain an odd number of generators of $ G/H$, with
negative parity.

So at the fixed point the non-abelian anomaly is restricted to 
the subgroup $H$ of the original group $G$.
But while the brane anomaly contains only $F^{\tilde a}_{\mu\nu} $ 
and reduces automatically to the anomaly 
of the unbroken subgroup $H$, in the bulk piece an additional
mixed term (\ref{mix-anomaly}) survives.

If we integrate over the compact space, we obtain two
contributions that affect the low energy effective 4d theory:
on one side part of the bulk anomaly survives and gives rise to 
derivative interactions between the zero modes and the Kaluza-Klein 
tower of the gauge field, on the other hand the localized piece 
reduces to the 4d anomaly of the zero modes, as in the case of the abelian
anomaly. 

Therefore, in order to have a viable 4d low energy theory, we need to 
impose the vanishing of the irreducible bulk anomaly and also require an 
anomaly-free configuration for the zero modes.

\section{An SO(10) GUT model}

We are now ready to consider a more interesting example, 
the SO(10) GUT model proposed in ref.~\cite{abc021}. 
We consider SO(10) Yang-Mills theory in 6d with N=1 supersymmetry. 
The gauge fields $A_M$ and the gauginos $\l_1$, $\l_2$ are 
conveniently grouped 
into vector and chiral multiplets of the unbroken N=1 supersymmetry in 4d,
\beq
A = (A_\m,\l_1)\;, \quad \S = (A_{5,6},\l_2)\;.
\eeq 
Here $A$ and $\S$ are matrices in the adjoint representation of SO(10).

Symmetry breaking is achieved by compactification on the orbifold 
$T^2/(Z_2^I\times Z_2^{PS}\times Z_2^{GG})$. 
The discrete symmetries $Z_2$ break the 
extended supersymmetry. They also break the SO(10) gauge group down to 
the subgroups SO(10), 
G$_{PS}$=SU(4)$\times$SU(2)$\times$SU(2), G$_{GG}$=SU(5)$\times$U(1)$_X$ 
and G$_{fl}$=SU(5)$'\times$U(1)$'$, at the four fixpoints 
$y_1 = y_{O} = (0,0)$, $y_2 = y_{PS}=(\p R_5/2,0)$, 
$y_3 = y_{GG}=(0,\p R_6/2)$ 
and $y_4 = y_{fl}=(\p R_5/2,\p R_6/2)$,
\bea\label{fixp}
P_IA(x,y_{O}-y)P_I^{-1} &=& \h_I A(x,y_{O}+y)\;,\\
P_{PS}A(x,y_{PS}-y)P_{PS}^{-1} &=& \h_{PS} A(x,y_{PS}+y)\;,\\
P_{GG}A(x,y_{GG}-y)P_{GG}^{-1} &=& \h_{GG} A(x,y_{GG}+y)\;,\\
P_{fl}A(x,y_{fl}-y)P_{fl}^{-1} &=& \h_{fl} A(x,y_{fl}+y)\;.
\eea 
Here $P_I=I$, the matrices $P_{PS}$ and $P_{GG}$ are given in the appendix,
and $P_{fl}=P_{GG}P_{PS}$, with $\h_{fl}=\h_{GG}\h_{PS}$. The parities are chosen 
as $\h_I=\h_{PS}=\h_{GG}=+1$. The extended supersymmetry is broken by choosing in 
the corresponding equations for $\S$  all parities $\h_i=-1$. 

\begin{figure}
\centering 
\includegraphics[scale=0.7]{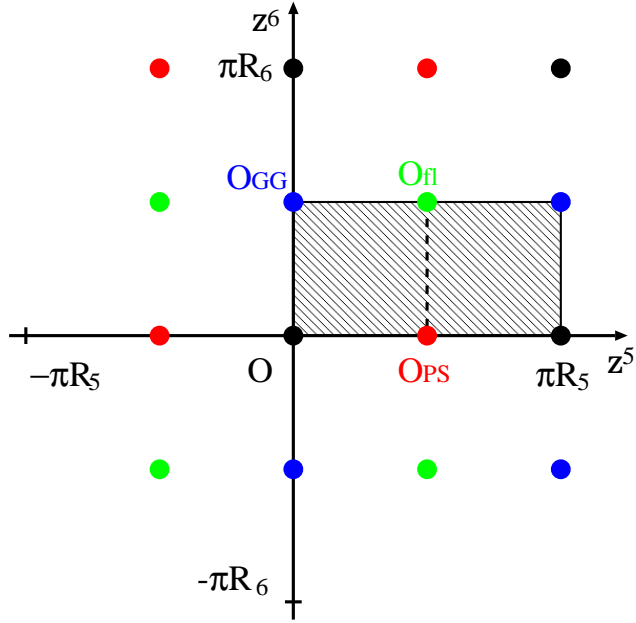}
\caption{Orbifold $T^2/(Z_2^I\times Z_2^{PS}\times Z_2^{GG})$ with the
fixpoints $O$, $O_{PS}$, $O_{GG}$, and $O_{fl}$.\label{fig:orb}}
\end{figure}

Figure~1 shows the four fixpoints, together with their three images each, on
the covering space $T^2$, with $z^5\in (-\p R_5,\p R_5]$ and $z^6\in (-\p R_6,\p R_6]$.
The physical region is obtained by folding the shaded region
along the dotted line and gluing the edges. The result is a `pillow' with the four 
fixpoints as corners. The unbroken gauge group of the effective 4d theory
is given by the intersection of the SO(10) subgroups at the fixpoints. In
this way one obtains the standard model group with an additional U(1) factor,
G$_{SM'}$= SU(3)$\times$SU(2)$\times$U(1)$_Y \times$U(1)$_X$. The zero modes of 
the vector multiplet $A$ form the gauge fields of G$_{SM'}$.

Matter and Higgs fields have been introduced motivated by the coset 
spaces E$_8$/(SO(10)$\times$H$_F$) where H$_F$ is a subgroup of SU(3)$\times$U(1) 
\cite{ong83}-\cite{ikk86}, which have previously been discussed in connection with
4d supersymmetric $\s$-models. In the case H$_F =$SU(3)$\times$U(1) the complex 
structure, and the corresponding representation of chiral multiplets is unique,
\beq\label{complex}
\O = ({\bf 16},{\bf 3})_1 + ({\bf 16^*},{\bf 1})_3 
+ ({\bf 10},{\bf 3^*})_2 + ({\bf 1},{\bf 3})_4\;.
\eeq
The SO(10) representations can in principle account for three quark-lepton generations, 
contained in the three {\bf 16}'s of SO(10), one mirror generation {\bf 16$^*$} and 
Higgs fields in the {\bf 10}'s. For bulk fields, however, only split multiplets
appear as zero modes in the effective 4d theory.

It is remarkable that the requirement of SO(10) bulk anomaly cancellation determines the
distribution of the SO(10) multiplets between bulk and branes. The vector multiplet 
is a {\bf 45}-plet of SO(10) which has a 6d anomaly. The irreducible anomalies
of fermions in the adjoint, vector and spinor representations are related
by (cf.~\cite{hmr02}),
\bea
a_{(4)}({\bf 45}) = 2 a_{(4)}({\bf 10})\;, \quad 
a_{(4)}({\bf 16}) = a_{(4)}({\bf 16^*}) = - a_{(4)}({\bf 10})\;.
\eea
Since fermions in vector and hypermultiplets have opposite chirality, the irreducible 
anomaly of the vector multiplet can be canceled by adding two {\bf 10}-plet 
hypermultiplets, $H_1$ and $H_2$. The complex structure (\ref{complex}) then
requires all three {\bf 10}-plets, and, consequently, also the {\bf 16$^*$}-plet to 
be bulk fields whereas the three {\bf 16}'s have to reside on branes. 

As discussed in \cite{abc021}, one can obtain the supersymmetric standard model with
right-handed neutrinos as effective 4d theory from this distribution of fields. A vacuum 
expectation value of {\bf 16$^*$} can then break $B-L$ and generate Majorana neutrino 
masses. To achieve this, the parities of the hypermultiplets have to be properly chosen,
\bea
P_I H(x,y_{O}-y) &=& \h_I H(x,y_{O}+y)\;, \label{so10a}  \\
P_{PS}H(x,y_{PS}-y) &=& \h_{PS} H(x,y_{PS}+y)\;,\label{so10b}\\
P_{GG}H(x,y_{GG}-y) &=& \h_{GG} H(x,y_{GG}+y)\label{so10c}\;,  
\eea
with $\h_i= \pm 1$ ($i=I,PS,GG$). The parities of the three {\bf 10}-plets $H_1$, $H_2$,
$H_3$ and the {\bf 16$^*$}-plet $\Phi^c$ are listed in table~\ref{tab:P10}.  
All hypermultiplets split 
under the extended 6d supersymmetry into two N=1 4d chiral multiplets, $H = (H,H')$. 
The two 4d left-handed fermions in the two chiral multiplets, $h_L$ and $h_L'$, transform
with respect to G as complex conjugates of each other. The 6d Weyl fermion is
$h=(h_L,h_L^{'c})$.
Invariance of the action requires that the parities of the 4d multiplets $H$ and $H'$
are opposite. We have denoted by $\h_i$ the parities of the first 4d chiral multiplet, 
and we have chosen $\h_I = +1$.

The discrete symmetry $Z_{PS}$ implies automatically a splitting
between the SU(2) doublets and the SU(3) triplets contained in the
{\bf 10}-plets.  The choice $\h_{PS} = +1$ leads to massless SU(2)
doublets and massive colour triplets (cf.~table~\ref{tab:P10}).
Choosing further $\h_{GG} = +1$ for $H_1$ and $\h_{GG} = -1$ for
$H_2$, selects the doublet $H^c$ from the SU(5) {\bf 5$^*$}-plet
contained in $H_1$, and the doublet $H$ from the SU(5) {\bf 5}-plet of
$H_2$ (cf.~table~\ref{tab:P10}). The doublets $H^c$ and $H$ have the
quantum numbers of the Higgs fields $H_d$ and $H_u$, respectively, in
the supersymmetric standard model.

\renewcommand{\arraystretch}{1.2}
\begin{table}[t]
  \begin{center}
   $\begin{array}[h]{|c||cc|cc|cc|cc|}\hline
     \mbox{SO(10)} &
     \multicolumn{8}{|c|}{ \mathbf{10} }
     \\ \hline
     G_{PS} &
     \multicolumn{2}{|c|}{ ( \mathbf{1}, \mathbf{2}, \mathbf{2}) } &
     \multicolumn{2}{|c|}{ ( \mathbf{1}, \mathbf{2}, \mathbf{2}) } &
     \multicolumn{2}{|c|}{ ( \mathbf{6}, \mathbf{1}, \mathbf{1}) } &
     \multicolumn{2}{|c|}{ ( \mathbf{6}, \mathbf{1}, \mathbf{1}) }
     \\ \hline
     G_{GG} &
     \multicolumn{2}{|c|}{ \mathbf{5}^\ast{}_{-2} } &
     \multicolumn{2}{|c|}{ \mathbf{5}{}_{+2} } &
     \multicolumn{2}{|c|}{ \mathbf{5}^\ast{}_{-2} }  &
     \multicolumn{2}{|c|}{ \mathbf{5}{}_{+2} }
     \\ \hline
        &  \multicolumn{2}{|c|}{H^c} & \multicolumn{2}{|c|}{H} &
     \multicolumn{2}{|c|}{G^c} & \multicolumn{2}{|c|}{G}
     \\
     {} &
    Z_2^{PS} & Z_2^{GG} &
    Z_2^{PS} & Z_2^{GG} &
    Z_2^{PS} & Z_2^{GG} &
    Z_2^{PS} & Z_2^{GG}
    \\ \hline \hline
     H_1 &
     + & + &
     + & - &
     - & + &
     - & -
     \\ \hline
     H_2 &
     + & - &
     + & + &
     - & - &
     - & +
     \\ \hline
     H_3 &
     - & + &
     - & - &
     + & + &
     + & -
     \\ \hline
    \end{array}$
  $\begin{array}[h]{|c||cc|cc|cc|cc|}\hline
    \mbox{SO(10)} & \multicolumn{8}{|c|}{ \mathbf{16}^\ast }
    \\ \hline
    G_{PS} &
    \multicolumn{2}{|c|}{ (\mathbf{4}^\ast, \mathbf{2}, \mathbf{1}) } &
    \multicolumn{2}{|c|}{ (\mathbf{4}^\ast, \mathbf{2}, \mathbf{1}) } &
    \multicolumn{2}{|c|}{ (\mathbf{4}, \mathbf{1}, \mathbf{2}) }  &
    \multicolumn{2}{|c|}{ (\mathbf{4}, \mathbf{1}, \mathbf{2}) }
    \\ \hline
    G_{GG} &
    \multicolumn{2}{|c|}{ \mathbf{10}^\ast{}_{+1} } &
    \multicolumn{2}{|c|}{ \mathbf{5}_{-3} } &
    \multicolumn{2}{|c|}{ \mathbf{10}^\ast{}_{ +1} } &
    \multicolumn{2}{|c|}{ \mathbf{5}_{-3},  \mathbf{1}{}_{+5} }
    \\ \hline
    {} &
    \multicolumn{2}{|c|}{ Q^c} &
    \multicolumn{2}{|c|}{L^c} &
    \multicolumn{2}{|c|}{U, E} &
    \multicolumn{2}{|c|}{D, N}
    \\
    {} &
    Z_2^{PS} & Z_2^{GG} &
    Z_2^{PS} & Z_2^{GG} &
    Z_2^{PS} & Z_2^{GG} &
    Z_2^{PS} & Z_2^{GG}
     \\ \hline
     \Phi^c &
     - & - &
     - & + &
     + & - &
     + & +
     \\ \hline
    \end{array}$
\caption{Parity assignment for the bulk $\mathbf{10}$ and $\mathbf{16}^\ast$ 
hypermultiplets. $H^c = H_d$ and $H = H_u$.}\label{tab:P10}
\end{center}
\end{table}

For the set of SO(10) fields given by eq.~(\ref{complex}) the irreducible 
bulk anomalies cancel, but reducible bulk anomalies remain.
In particular, the reducible anomaly of the {\bf 45} is not canceled by the 
anomalies of the
three {\bf 10}'s and the {\bf 16$^*$}, and the variation of the effective 
action reads
\bea
\d_v \G[A] = c \int {\rm tr}\(vdA\){\rm tr}\(F^2\)\; , 
\eea
where $c$ is a constant. 
This reducible anomaly can be canceled by the Green-Schwarz mechanism 
\cite{gs84},
where an antisymmetric tensor field $B$ with axion-like coupling is introduced,
\bea
S_B = c \int B {\rm tr}\(F^2\)\;.
\eea
Requiring $B$ to transform as 
\bea
\d_v B = - {\rm tr}\(vdA\)\;,
\eea
one obviously has $\d_v \G[A] + \d_v S_B = 0$.

In addition to the bulk anomalies one has to worry about the brane 
anomalies induced
at the four fixpoints by the chiral boundary conditions. 
Note that these anomalies contain also $F^2$ as do the reducible anomalies, 
but cannot be canceled by the Green-Schwarz mechanism since they contain
also information about the group index, absent in the case of the singlet
field $B$. 

In terms of the two 4d 
left-handed fermions contained in the chiral multiplets $H$ and $H'$ the 
left-handed
6d Weyl fermion is given by $h = (h_L, h_L^{'c})$. It transforms with respect 
to SO(10)
and its subgroups like $h_L$. The chiral boundary conditions 
(\ref{so10a})-(\ref{so10c}) together with the corresponding equations 
for $H'$ are then the analogue of the chiral boundary condition 
(\ref{chiralP1}), (\ref{chiralP2}) discussed in section~3. 
The SO(10) bulk symmetry  
is now broken to different subgroups at the four fixpoints. 
Correspondingly, bulk fields
split into representations of the common subgroup $G_{SM'}$. 

Consider as an example the {\bf 10}-plet $H_1$, with the parities listed in the table.
The split multiplets can be described by projection operators which act on the SO(10)
{\bf 10}-plet, i.e. $P_{H^c}$, $P_{H}$, $P_{G^c}$ and $P_{G}$. Different sums project
on representations of the fixpoint GUT groups, in obvious notation,
\bea
&&P_{H^c} +  P_{H} = P_{(1,2,2)}\;,\quad 
P_{G^c} +  P_{G} = P_{(6,1,1)}\;, \\
\label{defP1}
&&P_{H^c} +  P_{G^c} = P_{(5^*,-2)}\;,\quad 
P_{H} +  P_{G} = P_{(5,2)}\;, \\
\label{defP2}
&&P_{H^c} +  P_{G} = \tilde{P}_{(5^*,-2)}\;,\quad 
P_{H} +  P_{G^c} = \tilde{P}_{(5,2)}\; ,
\label{defP3}
\eea
where $\tilde{P}$ denote projection operators of flipped SU(5).

It is straightforward to calculate the nonabelian anomaly following the procedure
discussed in the previous section and generalizing to the presence of
three parities. The sum over modes now involves the projection
operators on all the states listed in the table as well as mode functions with the
corresponding parities. Instead of (\ref{Porbi6}) one obtains
\bea\label{Porbiso10}
&&\sum_{mn} e^{-{M_m^2+M_n^2\over \L^2}} 
\left\{ {1-\G_7\over 2}\left[\hat{P}_L \( P_{H^c}{\x_{+++}^{mn}}^2 +
P_{H}{\x_{++-}^{mn}}^2 + P_{G^c}{\x_{+-+}^{mn}}^2 + P_{G}{\x_{+--}^{mn}}^2\)
\right.\right.\NO\\
&&\hspace{2.3cm}\left.\left.
+\hat{P}_R \(P_{H^c}{\x_{---}^{mn}}^2 + P_{H}{\x_{--+}^{mn}}^2 
+ P_{G^c}{\x_{-+-}^{mn}}^2 + P_{G}{\x_{-++}^{mn}}^2\)\]\right.\NO\\
&&\hspace{0.7cm}\left.-{1+\G_7\over 2}\left[\hat{P}_R \( P_{H^c}{\x_{+++}^{mn}}^2 +
P_{H}{\x_{++-}^{mn}}^2 + P_{G^c}{\x_{+-+}^{mn}}^2 + P_{G}{\x_{+--}^{mn}}^2\)
\right.\right.\NO\\
&&\hspace{2.3cm}\left.\left.
+\hat{P}_L \(P_{H^c}{\x_{---}^{mn}}^2 + P_{H}{\x_{--+}^{mn}}^2 
+ P_{G^c}{\x_{-+-}^{mn}}^2 + P_{G}{\x_{-++}^{mn}}^2\)\]\right\}\;.
\eea
As in section~3 the various terms can be combined into two expressions which
yield the bulk and brane anomalies, respectively,
\bea
&&\sum_{mn} e^{-{M_m^2+M_n^2\over \L^2}} 
\left\{\G_7 {1\over 4} \sum_{bc}\({\x_{+bc}^{mn}}^2 + {\x_{-(-b)(-c)}^{mn}}^2\)\right. \NO\\ 
&&\hspace{1.3cm}\left.+ \(\hat{P}_R - \hat{P}_L\)
\(P_{H^c} \({\x_{+++}^{mn}}^2 - {\x_{---}^{mn}}^2\) +
P_{H} \({\x_{++-}^{mn}}^2 - {\x_{--+}^{mn}}^2\)\right.\right.  \NO\\
&&\hspace{3.6cm}\left.\left.
+ P_{G^c} \({\x_{+-+}^{mn}}^2 - {\x_{-+-}^{mn}}^2\) +
P_{G} \({\x_{+--}^{mn}}^2 - {\x_{-++}^{mn}}^2\)\)\right\}\; ;
\eea
here we have neglected a contribution to the bulk anomaly which vanishes in the limit
$\L \rightarrow \infty$.
Given the relations for sums over mode differences given in appendix C, 
one finally obtains for the anomaly,
\bea\label{an10-cov}
\hat{\ca}^a_{cov\,\,{\bf 10}} (x,y) &=& -\,
{1\over 8} {i^3\over (2\p)^3} {\rm tr}_{\bf 10}\(T^a F^3\) \NO\\
& & + {1\over 4}{i^2\over (2\p)^2}dz^5 dz^6 
\left[\;\d_O(y) {\rm tr}_{\bf 10} \(T^a F^2\) \right.\NO\\
& & \hspace{3cm} \left. 
+ \d_{PS}(y) {\rm tr}_{\bf 10}\((P_{(1,2,2)}-P_{(6,1,1)}) T^a F^2\)\right. \NO\\
& & \hspace{3cm} \left. 
+ \d_{GG}(y) {\rm tr}_{\bf 10}\((P_{(5^*,-2)}-P_{(5,2)}) T^a F^2\)\right. \NO\\
& & \hspace{3cm}\left.
+ \d_{fl}(y) {\rm tr}_{\bf 10}\((\tilde{P}_{(5^*,-2)}-\tilde{P}_{(5,2)}) T^a F^2\)\right]\;.
\eea
Going to the physical space $T^2/(Z_2^I\times Z_2^{PS}\times Z_2^{GG})$, the
bulk anomaly changes by a factor 8, whereas the fixpoint contributions only by
a factor 4 (cf. appendix~\ref{chap:appendixd}). The final result reads
\bea\label{an10}
\hat{\ca}^a_{\bf 10} (x,y) &=& 
 -\,{i^3\over (2\p)^3} {\rm tr}_{\bf 10}\(T^a F^3\) \NO\\
& & + \, {i^2\over (2\p)^2}dz^5 dz^6 
\left[\;\d_O(y) {\rm tr}_{\bf 10} \(T^a F^2\) \right.\NO\\
& & \hspace{3cm} \left. 
+ \d_{PS}(y) {\rm tr}_{\bf 10}\((P_{(1,2,2)}-P_{(6,1,1)}) T^a F^2\)\right. \NO\\
& & \hspace{3cm} \left. 
+ \d_{GG}(y) {\rm tr}_{\bf 10}\((P_{(5^*,-2)}-P_{(5,2)}) T^a F^2\)\right. \NO\\
& & \hspace{3cm}\left.
+ \d_{fl}(y) {\rm tr}_{\bf 10}\((\tilde{P}_{(5^*,-2)}-\tilde{P}_{(5,2)}) T^a F^2\)\right]\;.
\eea
At the fixpoints the SO(10) anomaly is reduced to an
anomaly of the unbroken subgroup, with a coefficient which is determined by the
difference of the anomalies into which the {\bf 10}-plet is split. Since SO(10)
is anomaly free in 4d, and also $(\mathbf{1},\mathbf{2},\mathbf{2})$ and 
$(\mathbf{6},\mathbf{1},\mathbf{1})$ have no $G_{PS}$ anomaly,
one is left with SU(5)$^2 \times$U(1)$_X$ and U(1)$_X^3$ anomalies at $y_{GG}$
and $ y_{fl}$. Using eqs.~(\ref{defP1})--(\ref{defP2}) one easily verifies that the
anomaly integrated over $T^2/Z_2^3$ equals the anomaly of the zero mode $H^c_1$.

It is now straightforward to write down the anomaly of the {\bf 16$^*$}-plet, given the
parities and split multiplets listed in the table,
\bea\label{an16}
\hat{\ca}^a_{\bf 16^*} (x,y) &=& 
-\, {i^3\over (2\p)^3} {\rm tr}_{\bf 16^*}\(T^a F^3\)\\
& & + {i^2\over (2\p)^2}\, dz^5 dz^6 \left[\d_0(y) 
{\rm tr}_{\bf 16}\(T^a F^2\) \right. \NO\\
&& \hspace{3cm}
+\left.\d_{PS}(y) {\rm tr}_{\bf 16^*}\((P_{(4,1,2)}-P_{(4^*,2,1)}) T^a F^2\)\right. \NO\\
&& \hspace{3cm}\left. + 
\d_{GG}(y) {\rm tr}_{\bf 16^*}\((P_{(5,-3)}+P_{(1,+5)}-P_{(10^*,1)}) T^a F^2\)\right. \NO\\
&& \hspace{3cm}\left. + 
\d_{fl}(y) {\rm tr}_{\bf 16^*}\((\tilde{P}_{(5,-3)}+\tilde{P}_{(1,+5)}
-\tilde{P}_{(10^*,1)}) T^a F^2\)\right]\;. \NO
\eea
Contrary to the {\bf 10}-plet anomaly, also on the PS fixpoint an anomaly is
generated.  The integrated anomaly equals again the sum of the contributions
from the zero modes $D$ and $N$. 

The {\bf 45}-plet of gauginos contributes to the bulk anomaly. At the
fixpoint $ y_{PS}$, it splits into $(\mathbf{15},\mathbf{1},\mathbf{1})$, 
$(\mathbf{1},\mathbf{3},\mathbf{1})$, $(\mathbf{1},\mathbf{1},\mathbf{3})$ and 
$(\mathbf{6},\mathbf{2},\mathbf{2})$, which are all anomaly free. At $y_{GG}$ and $y_{fl}$ the split 
multiplets are $\mathbf{24}_0$, $\mathbf{1}_0$,  $\mathbf{10}_{+4}$ and 
$\mathbf{10}^{\ast}_{-4}$; since $\mathbf{10}_{+4}$ and $\mathbf{10}^{\ast}_{-4}$ have
the same parities at these fixpoints \cite{abc01}, no anomaly is induced.


Summing all anomalies, of the {\bf 45}, the three {\bf 10}'s and the 
{\bf  16$^*$ },
the irreducible bulk anomalies cancel, and the reducible bulk anomaly can be 
canceled by the Green-Schwarz mechanism. 
There remain, however, brane anomalies
with contributions from the {\bf 10}-plet $H_3 $ and the {\bf 16$^*$}-plet 
$\Phi^c$,
\bea\label{anbrane}
\hat{\ca}^a_{brane} (x,y) &=&
{i^2\over (2\p)^2}dz^5 dz^6 \left\{\d_{PS}(y) {\rm tr}_{\bf 16^*}
\((P_{(4,1,2)}-P_{(4^*,2,1)}) T^a F^2\)\right. \\
& & \hspace{3cm}\left. + \d_{GG}(y) \[{\rm tr}_{\bf 10}
\( (P_{(5^*,-2)} - P_{(5,2)}) T^a F^2\) \right.\right. \NO\\
& & \hspace{4cm} \left. \left. + {\rm tr}_{\bf 16^*}
\((P_{(5,-3)} + P_{(1,+5)}- P_{(10^*,1)}) T^a F^2\)\]
\right. \NO\\
& & \hspace{3cm}\left. + 
\d_{fl}(y) \[{\rm tr}_{\bf 10}\((\tilde{P}_{(5^*,-2)}- \tilde{P}_{(5,2)}) T^a
F^2\) \right. \right. \NO\\ 
& & \hspace{4cm} \left.\left.
+ {\rm tr}_{\bf 16^*}\((\tilde{P}_{(5,-3)} + \tilde{P}_{(1,+5)}- 
\tilde{P}{(10^*,1)}) T^a F^2\)\] \right\}\;. \NO 
\eea
The result can be written in a simpler manner by noticing that
\bea
P_{PS} &=& P_{(4,1,2)}-P_{(4^*,2,1)}\;, \\
P_{GG} &=& P_{(5^*,-2)} - P_{(5,2)} = P_{(5,-3)} + P_{(1,+5)}- P_{(10^*,1)}\;, \\
P_{fl} &=& \tilde{P}_{(5^*,-2)}- \tilde{P}_{(5,2)} = 
\tilde{P}_{(5,-3)} + \tilde{P}_{(1,+5)} -\tilde{P}_{(10^*,1)}\;,
\eea
so we have for arbitrary matter content
\bea\label{anbrane2}
\hat{\ca}^a_{brane} (x,y) &=&
{i^2\over (2\p)^2}dz^5 dz^6  \sum_{all fields}
\left[\eta_{PS} \d_{PS}(y) {\rm tr} \(P_{PS} T^a F^2\)\right. \NO\\
&& \hspace{1cm}\left. + 
\eta_{GG} \d_{GG}(y) {\rm tr} \(P_{GG} T^a F^2\) + 
\eta_{PS} \eta_{GG} \d_{fl}(y) {\rm tr} \( P_{fl} T^a F^2\) \right]\;.
\eea
Hence the sign of the anomaly at the orbifold fixpoints depends on the
signs of the $\eta_i$. The full brane anomaly is given by a simple
trace containing the parity operators.  Note, that the brane anomalies
of the {\bf 10}-plets $H_1 $ and $H_2 $ cancel each other due to the
different values of $\eta_{GG}$.

It is important to realize that the conditions for vanishing brane anomalies 
are stronger than those requiring only the vanishing of the zero mode anomalies. 
This can be seen clearly from the formula above. Integrating over the compact
dimensions, we obtain
\bea\label{anbrane3}
\hat{\ca}^a_{brane} (x) &=&
\frac{1}{4}{i^2\over (2\p)^2} \sum_{all fields}
{\rm tr} \(\[\eta_{PS}P_{PS} + \eta_{GG}P_{GG} +\eta_{PS}\eta_{GG} P_{fl}\]
T^a F^2\)\; .
\eea
Clearly, the vanishing of the trace containing all parities does not
imply the vanishing of the single contributions in eq.~(\ref{anbrane2}). 

The cancellation of the brane anomalies (\ref{anbrane}) requires
additional degrees of freedom. One possibility is to add multiplets at
the fixpoints, whose contribution gives rise to a boundary term in
eq.~(\ref{abel-variation}).  In this case the matter content at each
brane has to be matched to cancel the corresponding anomaly.  A
simpler solution has been discussed in \cite{abc021}, the addition of
two more bulk fields: one {\bf 10}-plet, $H_4 $, and one {\bf
  16}-plet, $\Phi$.  Such a `partial doubling' is familiar from
supersymmetric $\s$-models~\cite{bl87}. In this case the irreducible
and reducible bulk anomalies as well as all brane anomalies cancel.
Note, that this choice of fields is still consistent with an eventual
embedding of all bulk and brane fields in to the {\bf 248} of $E_8 $
in 10d. Dimensional reduction of N=1 supersymmetry in 10d yields N=4
supersymmetry in 4d. Hence, the multiplicity of 4d chiral multiplets
with quantum numbers of the coset $E_8/(SO(10)\times H_F) $ has to be
less than or equal to four. In the model under consideration it would
be four for the bulk fields $H_{3,4} $ and $\Phi,\Phi^c$, two for the
bulk fields $H_1 $ and $H_2$, and one for the three {\bf 16}'s on the
brane.  The phenomenology of this model will be discussed elsewhere.

\section{Conclusions}

We have analyzed bulk and brane anomalies of 6d gauge theories compactified on
orbi\-folds. As in 5d theories, chiral boundary conditions at orbifold fixpoints
lead to brane anomalies in addition to the 6d bulk anomalies. 

For orbifold compactifications Fujikawa's method of calculating anomalies via 
the Jacobian of the path-integral measure is particularly well suited.
It yields the covariant anomaly as sum over mode functions of the chiral fermions.
Hence, boundary conditions at orbifold fixpoints, which project out some of the
modes, can be directly incorporated. For the discussion of anomaly cancellations 
the covariant anomaly is sufficient although it does not satisfy the
Wess-Zumino consistency conditions.

The main result of our analysis is very simple.  The bulk anomaly on
the orbifold equals the anomalies in flat space and on the torus.
Further, at a fixpoint with unbroken symmetry H, the non-abelian
anomaly of the bulk symmetry G reduces to an anomaly of H. If a bulk
multiplet of G is split into several multiplets of H at a fixpoint,
the H-anomaly is a sum of contributions of the split multiplets, with
signs which are determined by their parities.  The integrated anomaly equals
the anomaly of the zero modes.

For a given orbifold gauge model one can now easily determine all bulk and brane
anomalies whose cancellation strongly restricts allowed compactifications as well as
possible bulk and brane fields. In principle, it is straightforward to extend these 
results from six dimensions to eight and ten dimensions, and to include also 
gravitational anomalies.\\
 
\noindent
{\bf Acknowledgement}\\
\noindent
We would like to thank S.~Groot Nibbelink, A.~ Hebecker, H.~P.~Nilles, H.~B.~Nielsen,
E.~Poppitz and R.~Rattazzi for helpful discussions.

\section*{Appendices}
\begin{appendix}
\section{Conventions}
\label{chap:appendixa}
\ 
\vskip -0.5cm
\setcounter{equation}{0}
\renewcommand{\theequation}{A.\arabic{equation}}

In Minkowski space we shall work in the metric 
\begin{eqnarray}
  \eta_{MN}= \mbox{diag}(1,-1,-1,-1,-1,-1)~,
\end{eqnarray}
where $M,N = 0,1,2,3,5,6$. 

The $\G$-matrices in 6 dimensions, satisfying as usual
$\{\G_M,\G_N\}=2\eta_{MN}$, can be taken to be 
\bea
\G^{\m} = \(\begin{tabular}{cc} $\g^{\m}$&0 \\ 0&$\g^{\m}$ \end{tabular}\)\;,
\quad
\G^{5} = \(\begin{tabular}{cc} 0&$i\g_5$ \\ $i\g_5$&0 \end{tabular}\)\;,
\quad
\G^{6} = \(\begin{tabular}{cc} 0&$ -\g_5$ \\ $\g_5$&0 \end{tabular}\)\;,
\eea
with $\mu = 0,1,2,3$. Here $\gamma^\mu, \gamma^5$ are the 4d
$\gamma$-matrices in the notation of Itzykson-Zuber \cite{Itz-Zub}.
In particular we have
\bea
\gamma_5 = i \gamma^0 \gamma^1 \gamma^2 \gamma^3 \; ,
\eea
and 
\bea
\mbox{Tr} \left[\gamma_5 \gamma^\mu \gamma^\nu \gamma^\rho 
\gamma^\sigma\right] = - 4 i \epsilon^{\mu\nu\rho\sigma}\; ,
\eea
where we have chosen the convention $\epsilon^{0123} = +1$.

In 6d we define the analogous of $\gamma_5 $, $\G_7 (= \G^7)$, by
\begin{eqnarray}
  \G_7 =\G^0 \G^1 \G^2 \G^3 \G^5 \G^6 =
  \left( \begin{array}{cc} \g_5 & 0 \\ 0 & - \g_5 \end{array} \right)~.
\end{eqnarray}
Then, 
\begin{eqnarray}
  \mbox{Tr} \left[ \G_7 \G^M \G^N \G^O \G^P \G^Q \G^R \right]
      =  8 \epsilon^{MNOPQR}~,
\end{eqnarray}
where the antisymmetric tensor is chosen as $\epsilon^{012356} = +1$.
Note that in our conventions $\G^7 $ differs by a sign from that of~\cite{zwz84}.

To compute the change of the measure in the path integral, we perform
a Wick rotation and work in Euclidean space:
\begin{eqnarray}
  x^4 = i x^0~, ~~~~ \G^4 = i \G^0~, 
\end{eqnarray}
with the metric
\begin{eqnarray}
  \eta^E_{MN}= \mbox{diag}(-1,-1,-1,-1,-1,-1)= -\delta_{MN}~.
\end{eqnarray}
$\G_7$ and $\gamma_5$ are unchanged, {\it i.e.} we redefine them by
\begin{eqnarray}
  \G_7 &=& -i \G^4 \G^1 \G^2 \G^3 \G^5 \G^6 
  = i \G^1 \G^2 \G^3 \G^4 \G^5 \G^6 \; ,\\
\gamma^5 &=& \gamma^4 \gamma^1 \gamma^2 \gamma^3 = - \gamma^1 \gamma^2
  \gamma^3 \gamma^4\; .
\end{eqnarray}
Also the euclidean antisymmetric tensors are left unaffected, i.e.
\bea
\epsilon^{123456} &=& \epsilon^{123056} = - \epsilon^{012356} = - 1\; ,\\
\epsilon^{1234} &=& \epsilon^{1230} = - \epsilon^{0123} = - 1\; .
\eea
Then the traces over the euclidean $\gamma$-matrices are given by
\bea
\mbox{Tr} \left[\gamma_5 \gamma^\mu \gamma^\nu \gamma^\rho 
\gamma^\sigma\right] = + 4 \epsilon^{\mu\nu\rho\sigma}\; ,
\eea
and
\begin{eqnarray}
  \mbox{Tr} \left[ \G_7 \G^M \G^N \G^O \G^P \G^Q \G^R \right]
      = + 8 i \epsilon^{MNOPQR}~,
\end{eqnarray}
where the $\epsilon$-tensors carry euclidean indices.

The gauge fields of the euclidean Yang-Mills theory are introduced as
\begin{eqnarray}
  A_M = i A_M^a T^a ~,
\end{eqnarray}
where $T^a$ denote the hermitian generators of a Lie algebra.  The field strength
tensor is given by
\begin{eqnarray}
  F_{MN} = [ D_M, D_N]~,
\end{eqnarray}
with $D_M = \partial_M + A_M$.  Then, the kinetic term is 
\begin{eqnarray}
  {\cal L} = \frac{1}{4kg^2} \mbox{Tr}
  \left[ F_{MN} F^{MN} \right]~,
\end{eqnarray}
where $g$ is a gauge coupling and 
$\mbox{Tr}\left[T^aT^b\right] =k \delta^{ab}$.

In the text we present the anomaly in the euclidean space.
To obtain the usual expressions for the anomaly, 
note that the gauge field in the traditional notation and 
in Minkowski space is given by
\bea
{\cal F}_{MN} = - i F_{MN} \quad \mbox{and} \quad {\cal F}_{0M} = F_{4M}\; ,
\eea
where $M,N $ are spatial indices.  So we have
\bea
& & \e^{MNPQRS} F_{MN} F_{PQ} F_{RS} = i^2 \e^{MNPQRS}  {\cal F}_{MN}
{\cal F}_{PQ} {\cal F}_{RS}\\
& & \e^{\m\n\r\s} F_{\m\n} F_{\r\s} = i \e^{\m\n\r\s} {\cal F}_{\m\n} 
{\cal F}_{\r\s}\; .
\eea

\section{SO(10) matrices}
\setcounter{equation}{0}
\renewcommand{\theequation}{B.\arabic{equation}}

As well-known, the vector representation of SO(10) is given
by the $10\times 10$ real orthogonal matrices. Its
Lie algebra in the same representation corresponds to 
the antisymmetric $10\times 10$ real matrices. 
From these properties is then straightforward to realize that
the vector and the adjoint representations of SO(10) are always 
anomaly free in any dimension $2n$ with even $n$, since the trace of an 
odd number of generators vanishes exactly\footnote{ 
The spinor representation is also anomaly free apart in $d=8$
dimension.}. So, e.g. in 4d, SO(10) is usually regarded as
a safe group with respect to anomalies.

The traces of an even number of generators are non-vanishing.
For the case of four generators, giving the 6d bulk non-abelian 
anomaly, the normalization of the traces in the adjoint and 
spinor representation with respect to the vector representation
for SO(N) reads (cf. \cite{hmr02})
\bea
\mbox{ tr}_{\bf adj} F^4 &=& (N-8)\mbox{ tr}_{\bf vec} F^4 + 3 \(\mbox{ tr}_{\bf
  vec} F^2 \)^2 \;, \\
\mbox{ tr}_{\bf spin} F^4 &=& 
- 2^{(N-10)/2} \mbox{ tr}_ {\bf vec} F^4 + 3\, 2^{(N-14)/2}  \(\mbox{ tr}_{\bf vec} F^2 \)^2\;.
\eea
For the case of two generators, instead
\bea
\mbox{ tr}_{\bf adj} F^2 &=& (N-2)\mbox{ tr}_{\bf vec} F^2 \;, \\
\mbox{ tr}_{\bf spin} F^2 &=& 2^{(N-8)/2} \mbox{ tr}_ {\bf vec} F^2 \;.
\eea

Without loss of generality, we can take the group breaking 
parities in the vector representation to be
\bea
P_{PS} = 
\(\begin{array}{ccccc} -\sigma^0 &0&0&0&0\\
0&-\sigma^0 &0&0&0\\ 
0&0&-\sigma^0 &0&0\\
0&0&0&\sigma^0 &0\\
0&0&0&0&\sigma^0
\end{array}\)\;,\\
P_{GG} = \(
\begin{array}{ccccc} 
\sigma^2 &0&0&0&0\\
0&\sigma^2 &0&0&0\\ 
0&0&\sigma^2 &0&0\\
0&0&0&\sigma^2 &0\\
0&0&0&0&\sigma^2
\end{array}\)\;,
\eea
where $\sigma^0$ is the $2\times 2$ unity matrix, while $\sigma^2$ is
the Pauli matrix.
These operators belong to the involutive automorphisms of the 
Lie algebra of SO(10) and single out as invariant subalgebra 
the maximal compact subalgebras of the SO(10), i.e.
SO(6)$\times $ SO(4) and SU(5)$\times$U(1) respectively.
Note that $P_{PS}$ is a group element of SO(10) and therefore we have
also in this case, using $P_{PS}^{T} = P_{PS}$ and 
$P_{PS} T^a = T^a P_{PS} $,
\bea
{\rm tr} \left(P_{PS} T^a \{T^b, T^c\} \right) = 
- {\rm tr} \left(P_{PS} T^a \{T^b, T^c\} \right) = 0\; .
\eea
Therefore the anomaly on the Pati-Salam fixpoint is given only by 
the contribution of the spinor representation.

$P_{GG}$ and correspondingly $P_{fl} = P_{PS} P_{GG} $ are not SO(10) group 
elements and so a non-vanishing anomaly arises also from the vector
representation at $y_{GG}$ and $y_{fl}$.

\section{Mode functions on $T^2$}
\label{chap:appendixc}
\ 
\vskip -0.5cm
\setcounter{equation}{0}
\renewcommand{\theequation}{C.\arabic{equation}}

On the torus $T^2$ functions $\f(x,y)$, with $y=(z^5,z^6)$, can be expanded with 
respect to the following orthonormal basis, 
\bea
\f(x,y) = \sum_{m,n;a,b,c} \f^{mn}_{abc}(x) \x^{mn}_{abc}(y)\;.
\eea
Here $m,n$ are integers and $a,b,c = +,-$, with
\bea
& &\x_{+bc}^{mn}(y) = {1\over \sqrt{2\p^2 R_5 R_6 2^{\d_{m,0}\d_{n,0}}}} 
\cos\({m z^5\over R_5}+ {n z^6\over R_6}\)\; ,\\
& &\x_{-(-b)(-c)}^{mn}(y) = {1\over \sqrt{2\p^2 R_5 R_6}}
\sin\({m z^5\over R_5}+{n z^6\over R_6}\)\;;
\eea
$b(c)$ are $+$ or $-$ for $m(n)$ even or odd, respectively. 
The integers $m$ and $n$ run in the region 
$n \ge 0$ for $m=0$, and $\infty > n > - \infty$ for $m > 0$, for example.

Mode functions for all $m$ 
and $n$, even or odd, will be collectively denoted by $\x_{\pm}^{mn}$. The two sets of 
mode functions, $\x_+$ and $\x_-$, are related by differentiation,
\bea
\partial_5 \x^{mn}_{+bc} &=& - M_m\x^{mn}_{-bc}\;, \quad M_m = {m\over R_5}\;,\\
\partial_5 \x^{mn}_{-bc} &=& + M_m\x^{mn}_{+bc}\;,\\
\partial_6 \x^{mn}_{+bc} &=& - M_n\x^{mn}_{-bc}\;, \quad M_n = {n\over R_6}\;,\\
\partial_5 \x^{mn}_{-bc} &=& + M_n\x^{mn}_{+bc}\;,
\eea
and satisfy the orthonormality conditions
\bea
\int_{-\p R_5}^{\p R_5}dz^5\int_{-\p R_6}^{\p R_6}dz^6\ 
\x^{mn}_{abc}(y)\ \x^{m'n'}_{a'b'c'}(y) = \d_{mm'}\d_{nn'}\d_{aa'}\d_{bb'}\d_{cc'}\;.
\eea

The mode functions are even/odd under reflections at the four fixpoints of the
orbifold $T^2/(Z_2^I\times Z_2^{PS}\times Z_2^{GG})$, 
$y_1 \equiv y_{O} = (0,0)$, $y_2 \equiv y_{PS} = (\p R_5/2,0)$, 
$y_3 \equiv y_{GG} = (0, \p R_6/2)$, $y_4 \equiv y_{fl} = (\p R_5/2,\p R_6/2)$,
\bea
\x^{mn}_{\pm bc}(-y) &=& \pm\x^{mn}_{\pm bc}(y)\; ,\\
\x^{mn}_{a\pm c}(y_2-y) &=& \pm\x^{mn}_{a\pm c}(y_2+y)\; ,\\
\x^{mn}_{ab\pm}(y_3-y) &=& \pm\x^{mn}_{ab\pm}(y_3+y)\; ,\\
\x^{mn}_{a\pm\pm}(y_4-y) &=& \pm\x^{mn}_{a\pm\pm}(y_4+y)\; .
\eea
Furthermore, the following completeness relations hold,
\bea
\sum_{mn}\left(\x^{mn2}_{+++}(y)-\x^{mn2}_{---}(y)\right) &=& \d_{++}(y)\;,\\
\sum_{mn}\left(\x^{mn2}_{++-}(y)-\x^{mn2}_{--+}(y)\right) &=& \d_{+-}(y)\;,\\
\sum_{mn}\left(\x^{mn2}_{+-+}(y)-\x^{mn2}_{-+-}(y)\right) &=& \d_{-+}(y)\;,\\
\sum_{mn}\left(\x^{mn2}_{+--}(y)-\x^{mn2}_{-++}(y)\right) &=& \d_{--}(y)\;,
\eea
where
\bea
\d_{++}(y) &=& {1\over 4}\left(\d_O(y) + \d_{PS}(y) + \d_{GG}(y) + \d_{fl}(y)\right)\;,\\
\d_{+-}(y) &=& {1\over 4}\left(\d_O(y) + \d_{PS}(y) - \d_{GG}(y) - \d_{fl}(y)\right)\;,\\
\d_{-+}(y) &=& {1\over 4}\left(\d_O(y) - \d_{PS}(y) + \d_{GG}(y) - \d_{fl}(y)\right)\;,\\
\d_{--}(y) &=& {1\over 4}\left(\d_O(y) - \d_{PS}(y) - \d_{GG}(y) + \d_{fl}(y)\right)\;,
\eea
with
\bea
\d_{O}(y) &=& {1\over 4}\(\d(y+y_1)+\d(y+y_1-2y_2)\right.\NO\\
&&\hspace{0.5cm}\left.+\d(y+y_1-2y_3)+\d(y+y_1-2y_4)\)\;,\\
\d_{PS}(y) &=& {1\over 4}\(\d(y+y_2)+\d(y+y_2-2y_2)\right.\NO\\
&&\hspace{0.5cm}\left.+\d(y+y_2-2y_3)+\d(y+y_2-2y_4)\)\;,\\
\d_{GG}(y) &=& {1\over 4}\(\d(y+y_3)+\d(y+y_3-2y_2)\right.\NO\\
&&\hspace{0.5cm}\left.+\d(y+y_3-2y_3)+\d(y+y_3-2y_4)\)\;,\\
\d_{fl}(y) &=& {1\over 4}\(\d(y+y_4)+\d(y+y_4-2y_2)\right.\NO\\
&&\hspace{0.5cm}\left.+\d(y+y_4-2y_3)+\d(y+y_4-2y_4)\)\;.
\eea
Summing over all even and odd modes yields
\bea\label{com6}
\sum_{mn}\left(\x^{mn2}_{+}(y)-\x^{mn2}_{-}(y)\right) &=& 
\sum_{bc}\sum_{mn}\left(\x^{mn2}_{+bc}(y)-\x^{mn2}_{-(-b)(-c)}(y)\right) \NO\\
&=& \d_{++}(y) + \d_{+-}(y) + \d_{-+}(y) + \d_{--}(y) \NO\\ 
&=& \d_O(y)\;.
\eea

A complete set of orthonormal modes $\x^m_{\pm b}$ on the circle $S^1$ is obtained 
by dimensional reduction,
\begin{eqnarray}
  \x^{m}_{\pm b}(z^5) \equiv
  \sqrt{2\pi R_6} ~\xi_{\pm b c}^{m0} (y)~.
\end{eqnarray}
The corresponding orthonormality and completeness relations are ($y=z^5$),
\bea
\int_{-\p R_5}^{\p R_5}dy\ 
\x^{m}_{ab}(y)\ \x^{m'}_{a'b'}(y) = \d_{mm'}\d_{aa'}\d_{bb'}\;,
\eea
\bea\label{com5}
\sum_{m}\left(\x^{m2}_{+}(y)-\x^{m2}_{-}(y)\right) &=& 
\sum_{b}\sum_{m}\left(\x^{m2}_{+b}(y)-\x^{m2}_{-b}(y)\right) \NO\\ 
&=& {1\over 2} \left(\d(y) + \d(y-\p R_5)\right) \;.
\eea

\section{Physical versus covering space anomalies}
\setcounter{equation}{0}
\renewcommand{\theequation}{D.\arabic{equation}}
\label{chap:appendixd}

\subsection{$T^2/Z_2$}

The physical space of the orbifold $T^2/Z_2$
can be parameterize by the rectangle 
$ \((-\pi R_5, \pi R_5], [0, \pi R_6]\)$, while the covering
space is given by the torus, i.e. 
$ \((-\pi R_5, \pi R_5],(-\pi R_6, \pi R_6]\)$.
Let us extend a smooth function $f$ on the orbifold to 
the whole covering space using the orbifold symmetry,
keeping 
\bea
\int_{T^2/Z_2}\! d^2y\, f(y)\, =
\int_{T^2}\! d^2y\, f_{cov} (y)\; .
\eea
It is then easy to see that we have
\bea
f_{cov} (z^5,z^6) = \left\{
\begin{array}{ll}
{1\over 2} f(z^5,z^6)\;, & \hspace{1cm} z^6 \geq 0 \\
{1\over 2} f(z^5,- z^6)\;, & \hspace{1cm} z^6 < 0
\end{array} \right. 
\eea

Note on the other hand that both spaces contain fully 
the same fixed points,
i.e. $y_1 = (0,0)$, $y_2 = (\p R_5,0)$, $y_3 = (0,\p R_6)$ and 
$y_4 = (\p R_5,\p R_6)$. For a localized delta-function at any
fixpoint $y_i$ we have therefore automatically
\bea
\int_{T^2/Z_2}\! d^2y\, \delta(y-y_i)\; =  
\int_{T^2}\! d^2y\, \delta(y-y_i)\; .
\eea
So for a generic covering function
\bea
\ca_{cov} (y) = f_{cov} (y) + \delta(y-y_i) \; ,
\eea
the physical function on the orbifold $y \in T^2/Z_2 $ reads simply
\bea
\ca (y) = 2 f_{cov} (y) + \delta(y-y_i)\; .
\eea

\subsection{$T^2/(Z_2\times Z_2\times Z_2)$}

The physical space of the orbifold $T^2/(Z_2\times Z_2\times Z_2)$
can be parameterized by the rectangle 
$\( [0, \pi R_5), [0, \pi R_6/2]\)$, while the covering
space is given again by the torus, i.e. 
$\( (-\pi R_5, \pi R_5],(-\pi R_6, \pi R_6]\)$.
The volume of the torus is eight times the volume of the orbifold 
$T^2/(Z_2\times Z_2\times Z_2)$. 
So for any smooth function respecting the orbifold symmetry, 
we can again define
\bea
\int_{T^2/Z_2^3}\! d^2y\, f (y)\; =  
\int_{T^2}\! d^2y\, f_{cov} (y)\; .
\eea
Then the function on the covering space, satisfying the above
relation, is given by
\bea
f_{cov} (y) = \left\{
\begin{array}{ll}
{1\over 8} f(y) \;, & \hspace{1cm} y \in T^2/Z_2^3 \\
{1\over 8} f(P(y)) \;, & \hspace{1cm} y \notin T^2/Z_2^3, P(y) \in T^2/Z_2^3\\  
\end{array} \right. 
\eea
where $P$ is the action of the orbifold parities that brings 
$y$ from the torus inside the physical space.

Note on the other hand that the torus contains four times more
fixpoints than the orbifold physical space, as shown in fig.~1. 
Then for a localized function on a fixpoint, we have 
for $i=O, PS, GG, fl$ (cf. appendix C)
\bea
\int_{T^2/Z^3_2}\! d^2y\, \delta_i (y)\; =  
{1\over 4} \int_{T^2}\! d^2y\, \delta_i (y)\; .
\eea
So, generically, for a covering function on the torus given by
\bea
\ca_{cov} (y) = f_{cov} (y) + \delta_i (y) \; ,
\eea 
we obtain on the orbifold $T^2/(Z_2\times Z_2\times Z_2)$
the physical function
\bea
\ca (y) = 8 f_{cov} (y) + 4  \delta_i (y)\; .
\eea

\end{appendix}

\end{document}